\documentclass[10pt,conference,letter]{IEEEtran}
\usepackage{array,color,graphicx,amsmath,amssymb,amsthm,epsfig,mathrsfs,epstopdf}
\usepackage{amsthm}

\usepackage{multicol}

\begin{document}

\title{Cross-Layer Strategies for Throughput Maximization in a Data Aggregating Wireless Network}

\author{
\IEEEauthorblockN{Easwar Vivek Mangipudi}
\IEEEauthorblockA{Department of Electrical Engineering\\
Indian Institute of Technology Madras\\
Chennai 600036, India.\\
Email: easwar.vivek@gmail.com}
\and
\IEEEauthorblockN{Venkatesh Ramaiyan}
\IEEEauthorblockA{Department of Electrical Engineering\\
Indian Institute of Technology Madras\\
Chennai 600036, India.\\
Email: rvenkat@ee.iitm.ac.in}
\and
\IEEEauthorblockN{Srikrishna Bhashyam}
\IEEEauthorblockA{Department of Electrical Engineering\\
Indian Institute of Technology Madras\\
Chennai 600036, India.\\
Email: skrishna@ee.iitm.ac.in}
}

\IEEEoverridecommandlockouts
\IEEEpubid{\makebox[\columnwidth]{978-1-4673-5952-8/13/\$31.00~\copyright~2013 IEEE \hfill} \hspace{\columnsep}\makebox[\columnwidth]{ }}
\maketitle

\begin{abstract}
We consider an ad hoc wireless network where all nodes have data to send to a single destination node called the sink. We consider a linear placement of the wireless nodes with the sink at one end. We assume that the wireless nodes transfer data to the sink using single hop direct transmission and that the nodes are scheduled one at a time by a central scheduler (possibly the sink).
In this setup, we assume that the wireless nodes are power limited and our network objective (notion of fairness) is to maximize the minimum throughput of a node subject to the individual node power constraints.
In this work, we consider network designs that permit different node transmission time, node transmission power and node placements, and study cross-layer strategies that seek to maximize the minimum node throughput.
Using simulations, we characterize the performance of the different strategies and comment on their applicability for various network scenarios.
\end{abstract}

\section{Introduction}
We consider an ad hoc wireless network where all the wireless nodes have data to send
to a common destination node called the sink.
We assume that the wireless nodes are placed
along a straight line with the sink at one end.
This could be a sensor network where the sensor nodes, placed regularly along a line, generate
data and send it to a common fusion center or the sink.
In an ad hoc network deployment
for an emergency scenario, the sink could be the Gateway to the Internet and the wireless nodes could correspond to access points or base stations set up temporarily along a
highway. The ad hoc network of the access points would provide
the necessary backbone to establish a communication setup for search and rescue operations.
The network objective in such a setup would be to provide fair throughput for every wireless node and
to provide bounded delay for real time applications.

In this work, we assume that the wireless nodes transfer data to the sink directly in a single hop and that the wireless nodes are scheduled one at a time by a central scheduler, i.e., there is no multihopping or spatial reuse. Distributed medium access protocols (e.g., IEEE 802.11 DCF) are usually popular in such network scenarios, however, they do not guarantee QoS required for crucial real time applications. Also, multihopping with spatial reuse requires coordinating simultaneous transmissions which incurs communication overheads making it not so useful when there are only a few tens of wireless nodes in the network (typical of an emergency network scenario). The centralized scheduling strategy and the single hop assumption ensures that the channel access delay can be bounded and a guaranteed QoS can be provided to all the wireless nodes.  In an emergency network scenario, we envision that the wireless nodes/access points would communicate in a TDM fashion with the sink, thus, ensuring minimum throughput and bounded delay with the sink; the access point could use popular distributed MAC protocols like WiFi or ZigBee to provide access to the end users.

We assume that the wireless nodes are individually power constrained (except the sink node) and our network objective in this setup is to maximize the minimum throughput of the wireless nodes subject to the individual node power constraint. For the uplink traffic scenario studied in this work, clearly, the nodes that are located close to the sink have a better channel to the sink than the nodes that are placed farther from the sink. In this work, we consider network designs that permit different node transmission time, node transmission power and node placement, and study cross-layer strategies that seek to optimize the network performance. Using simulations, we characterize the performance of the different strategies and comment on their applicability for various network scenario.

\subsection{Related Literature}
The seminal work by Gupta and Kumar \cite{GK2000} describes the max-min fair
capacity of a large ad hoc wireless network with random source-destination pairs.
The asymptotic transport capacity of a many-to-one data gathering wireless channel was studied
in \cite{HEG05}.
In our work, we consider arbitrary placement of a fixed number of nodes and
permit using different node transmission time and transmission power to maximize the network throughput.
In \cite{STAG03}, Toumpis and Goldsmith, have characterized the rate region of a wireless network for an arbitrary but fixed placement of nodes but they do not optimize on the network parameters.

For a regular linear array of nodes, Giridhar and Kumar \cite{AGPRK05}, have studied the functional lifetime of the network as a function of the initial node energies. In \cite{LEELA10}, Leela has studied the energy optimal routing strategy for a linear array of data aggregating wireless nodes. In our work, we aim to maximize the minimum throughput of any node for an average power constraint. Linear placement of nodes has been studied in \cite{Chattopadhyay}, \cite{AA10} and \cite{AA09} as well though in a context of relay placement.

In \cite{QB04}, Qin and Berry propose an opportunistic scheduling scheme called opportunistic splitting algorithm for a single hop, fading wireless channel. In \cite{HBK09}, Hou et al. characterize the rate region for a single hop, fading wireless channel and discuss general network utility optimal scheduling strategies as well. We do not consider fading channels, however, we expect that considering scenarios like fading and general network utility maximisation will only add to the work reported in this paper.

\subsection{Outline}
In section~\ref{sec:network_model}, we describe the network model and in section~\ref{sec:network_objective}, we describe the network objective and the optimization framework. In section~\ref{sec:numerical_work}, we report the performance of the wireless network for the different operating strategies and in section~\ref{sec:conclusion}, we conclude the work and discuss future work.

\section{Network Model}
\label{sec:network_model}
\begin{figure}
\begin{center}
\includegraphics[scale=0.2]{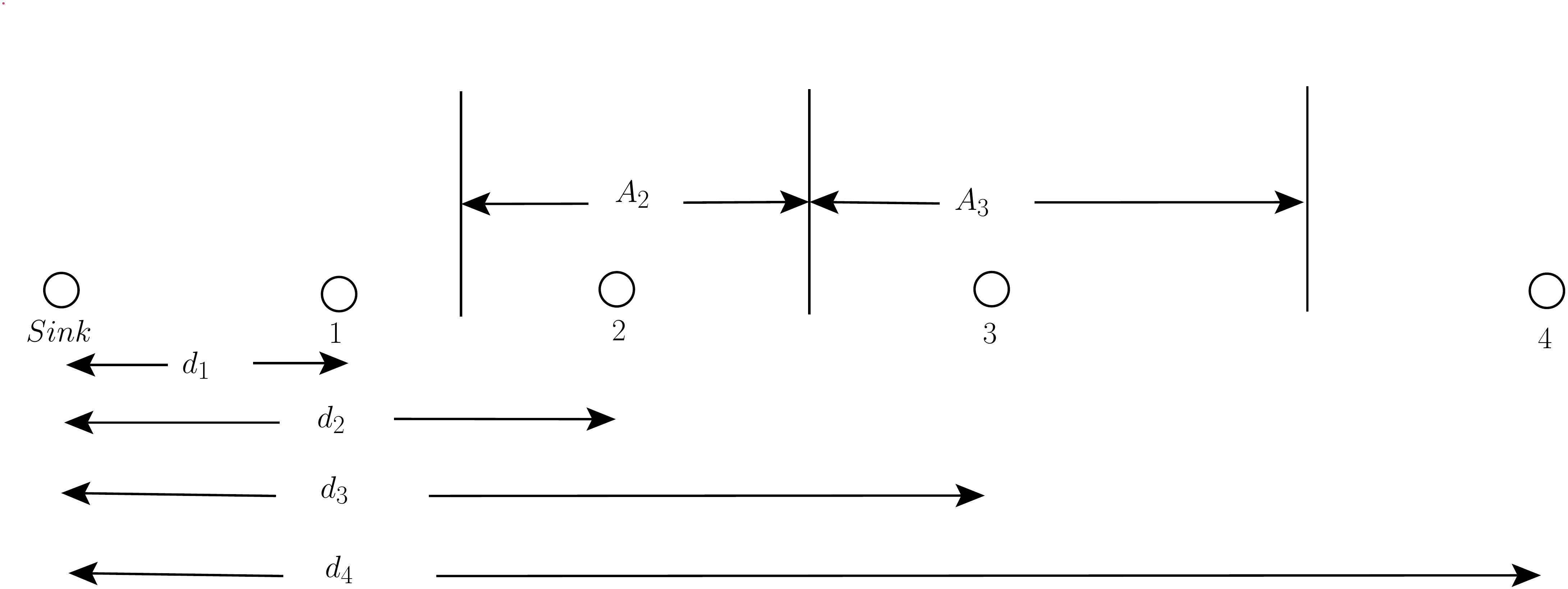}
\caption{A linear arrangement of nodes. All the nodes send data to the sink by single hop direct communication.}
\label{fig:figure1}
\end{center}
\end{figure}
We consider a wireless network with a linear arrangement of nodes as shown in Figure~\ref{fig:figure1}.
The sink is assumed to be located at one end of the network and the nodes, numbered, $1,2,\cdots,n$,
are located at a distance $d_1, d_2, \cdots, d_n$ from the sink, where $0 \leq d_1 \leq d_2 \leq \cdots \leq d_n$.
The wireless channel gain between a node $i$ and the sink is assumed to be a constant and is modeled as a function of the path loss only. In this work, we assume that the fixed path loss between a node $i$ and the sink is
$\frac{1}{d_i^{\eta}}$, where $\eta$ is the path loss exponent ($\eta \geq 2$).

A central scheduler schedules the wireless nodes one at a time. We consider
a frame of size $T$ seconds such that every node is scheduled once in a frame.
The frame duration $T$ bounds the maximum delay of access to the sink for any node $i$.
Let $t_1, t_2, \cdots, t_n$ be the amount of time allocated to the users in a frame,
where $0 \leq t_1, \cdots, t_n \leq T$ and $\sum_{i=1}^n t_i \leq T$.

We assume that the
nodes have an average power constraint of $P_i$ with the total network power being $\bar{P}_n = \sum_{i=1}^n P_i$.
The average energy spent by a node in a frame is $T P_i$.
When a node $i$ is scheduled in a frame for a duration of $t_i$ seconds with an average energy of $T P_i$,
 then, we assume that the node can achieve a Shannon capacity of $C_i(t_i, P_i, d_i) := W t_i \log\left(1 + \frac{\frac{T P_i}{t_i}}{\sigma^2 d_i^\eta} \right)$; $W$ is the system bandwidth and $\sigma^2$ is the average noise power.
The capacity expression does not include a term for interference in communication as we assume that the nodes are scheduled one at a time (no spatial reuse).

In this setup, the network objective is to maximize the minimum throughput of the wireless nodes, i.e.,
\[ \max_{
  \Bigg\{ {\substack{\sum_{i=0}^n t_i \leq T \\ \sum_{i=0}^n P_i \leq \bar{P_n} \\ d_1 \leq \cdots \leq d_n}}  \Bigg\} } \min \Big( C_1(t_1, P_1, d_1), \cdots, C_n(t_n, P_n, d_n) \Big) \]
subject to the time and power constraints and node placements.
The nodes that are closer to the sink have a better channel to the sink relative to the nodes that are farther away.
An equal share of the network resources would severely hurt the farther nodes and bring down the network performance.
In this work, we study cross-layer strategies that seek to maximize the network throughput by optimizing the node transmission
time, node transmission power and also the node position.
In Section~\ref{sec:network_objective} we describe in detail the network objective and the optimization problem and in Section~\ref{sec:numerical_work},
we report the network performance.
Without loss of generality, we will take $T = 1, W = 1$ and $\sigma^2 = 1$. Then, the time constraint reduces to
$\sum_{i=1}^n t_i \leq 1$ and the achievable capacity becomes $C_i(t_i, P_i, d_i) = t_i \log\left( 1 + \frac{\frac{P_i}{t_i}}{d_i^\eta} \right)$.

\section{Network Optimization}
\label{sec:network_objective}
\subsection{Basic Reference Model}
Consider a network arrangement where the nodes are placed at regular intervals with equal distance between adjacent nodes,
i.e., $d_{i+1} = (i+1) d_1$ for all $i \geq 1$ and $d_1 > 0$. Also, suppose that the individual node power constraint
$P_i = P$ for all $i$ (and the total network power $\bar{P}_n = n P$) and the fraction of time allocated to a node is $t_i = \frac{1}{n}$ for all $i$.
The network model corresponds to a basic scenario where all the nodes have equal access to the channel, energy and traffic. The achievable capacity for a node in this setup is
$C^{BR}_i = \frac{1}{n} \log\left(1 + \frac{n P}{(i d_1)^{\eta}} \right)$ and the max-min capacity of the network is $C^{BR} = C^{BR}_n = \frac{1}{n} \log\left(1 + \frac{n P}{d_1^{\eta}} \frac{1}{n^{\eta}} \right)$.
In this work, we consider the above network performance as the basic reference model. In the following subsections, we will describe various network design strategies that adapt the node transmission time, node average power and the node location to maximize the network throughput. We will use the basic reference model to compare and analyze the performance of the various strategies.

\subsection{Time Adaptation}
The time adaptation scheme permits nodes to use different transmission time $t_i$ subject to the overall frame constraint, $\sum_{i=1}^n t_i \leq 1$. The nodes are placed at regular intervals such that $d_{i+1} = (i+1) d_1$ for all $i \geq 1$ and $d_1 > 0$, and the individual node power constraint $P_i = P$ for all $i$. The achievable capacity in this setup for any node $i$ is $C^{TA}_i(t_i) = t_i \log\left(1 + \frac{\frac{P}{t_i}}{(i d_i)^{\eta}}\right)$ and the max-min capacity of the network is
\[ C^{TA} = \max_{\{ t_i : t_i \geq 0, \sum_{i=1}^n t_i \leq 1\} } \min \left(C^{TA}_1, \cdots, C^{TA}_n \right) \]
We note that the time adaptation scheme can easily be implemented by the central scheduler that schedules individual users.

\subsubsection*{Analysis}
Let $d_1 = 1$. Let $\{t_i^*\}$ be the time allocation that optimizes the max-min throughput $C^{TA}$.
Then for $1 \leq i,j \leq n$, we have,
\[t_i^* \log\left(1+\frac{\frac{P}{t_i^*}}{i^{\eta}}\right)=t_j^* \log\left(1+\frac{\frac{P}{t_j^*}}{j^{\eta}}\right)\] From the above expression, we see that $t_1^* \leq t_2^* \cdots \leq t_n^*$.
This is expected because at the optimal solution, the nodes that are farther away are allocated more transmission time to compensate for their poor channel gain.

The max-min throughput $C^{TA}$ is upper bounded by $\bar{C}^{TA}=\log(1+\frac{P}{n^\eta})$, which is the throughput obtained by the last node (node $n$) when all the time is allotted to it. The ratio of $\bar{C}^{TA}$ to $C^{BR}$ then upper bounds the performance of time adaptation scheme in comparison with the base reference model. For large $n$, we have,
\begin{equation} \label{eq:TAlimit}
\lim_{n \rightarrow \infty} \frac{\bar{C}^{TA}}{C^{BR}}=\lim_{n \rightarrow \infty} \frac{\log(1+\frac{P}{n^{\eta}})}{\frac{1}{n}*\log(1+\frac{P}{n^{\eta-1}})}=1
\end{equation}
i.e., for large $n$, the time adaptation scheme does not provide any gain over the base reference model. In section~\ref{sec:numerical_work}, from simulations, we observe that the time adaptation provides substantial gain for small $n$.

\subsection{Power Adaptation}
The power adaptation scheme permits nodes to have different average node power $P_i$ subject to the overall network power constraint, $\sum_{i=1}^n P_i \leq \bar{P}_n$. The nodes are placed at regular intervals such that $d_{i+1} = (i+1) d_1$ for all $i \geq 1$ and $d_1 > 0$, and the node have equal transmission times, $t_i = \frac{1}{n}$. The achievable capacity in this setup for any node $i$ is given by $C^{PA}_i(P_i) = \frac{1}{n} \log\left(1 + \frac{n P_i}{(i d_i)^{\eta}}\right)$ and the max-min capacity of the network is
\begin{equation} \label{eq:PA}
 C^{PA} = \max_{\{ P_i : P_i \geq 0, \sum_{i=1}^n P_i \leq \bar{P}_n\} } \min \Big(C^{PA}_1, \cdots, C^{PA}_n \Big)
\end{equation}
Unlike the time adaptation scheme, the power adaptation scheme requires the network to be designed apriori and the nodes have to be provisioned before deployment. Power adaptation is possible by provisioning different capacities of battery (or, say, solar cells) in different nodes.

\subsection*{Analysis}
Let $d_1 = 1$. Let $\{P_i^*\}$ be the power allocation that optimizes the max-min throughput $C^{PA}$.
Then for $1 \leq i,j \leq n$, we have,
\[\frac{1}{n} \log\left(1+\frac{n P_i^*}{i^{\eta}}\right)= \frac{1}{n} \log\left(1+\frac{ n P_j^*}{j^{\eta}}\right)\]
From the above expression, we see that $\frac{P_i^*}{i^{\eta}} = \frac{P_j^*}{j^{\eta}}$ or
$P_i^* = i^{\eta} P_1^*$. Using the network power constraint $\sum_{i=1}^n P_i^* \leq \bar{P}_n$, we get $P_i^* = i^{\eta} \frac{\bar{P}_n}{\sum_{i=1}^n i^{\eta}}$.
Clearly, $P_1^* \leq P_2^* \cdots \leq P_n^*$.
Here again, the nodes that are farther away are allocated more transmission power to compensate for their poor channel gain.

For $\eta = 2$ and for $\bar{P}_n = n P$, we have, for $1 \leq i \leq n$,
$P_i^* = i^{2} \frac{n P}{\sum_{j=1}^n j^2} = \frac{6 n i^2 P}{n (n+1) (2n + 1)}$.
Substituting for the optimal power allocation in the throughput expression, the ratio of the max-min throughput with power adaptation and with the base reference model is given by, for large $n$,
\[   \frac{\log(1+\frac{6npn^2}{(n+1)(2n+1)n^2} )  }{\log (1+\frac{nP}{n^2}) } \approx 3 \]
The power adaptation scheme provides a bounded performance with respect to the base reference model for $\eta = 2$. However, for $\eta > 2$, in section~\ref{sec:numerical_work}, we note that power adaptation scheme provides significant gains in comparison with the base reference model.

\subsection{Node Placements}
In this scheme, we consider arbitrary node placements to optimize the network performance. In the earlier schemes (time adaptation and power adaptation), we had assumed that the nodes are regularly spaced. The regular placement of the wireless access points would imply that the access points would cater to similar sized regions and hence, we aimed to maximize the minimum throughput of the nodes. In this section, we permit arbitrary placements; this would imply that the access points can now cater to regions of different sizes depending on the inter-node placements. 
Consider an arbitrary location of nodes $\{ d_i \}$ such that $0 \leq d_1 \leq \cdots \leq d_n \leq n$.
For fixed node transmission times, $t_i = \frac{1}{n}$ and fixed node power constraints, $P_i = \frac{\bar{P}_n}{n} = P$, the achievable capacity of a node $i$ is given by
$C^{NP}_i(d_i) = \frac{1}{n} \log\left(1 + \frac{n P}{d_i^{\eta}} \right)$.
We consider the following network objective for this scenario,
\[ C^{NP} = \max_{\{ \{d_i\} : 0 \leq d_1 \leq \cdots \leq d_n \leq n \}} \min_{i} \left( \frac{C^{NP}_i}{A_i} \right) \]
where $A_i$ is the network area serviced by node $i$ corresponding to the placements $\{ d_i \}$.
The metric of interest is the throughput per unit region serviced by the wireless node. In this work, we compute the service region $A_i$ using a Voronoi tessellation, as shown in Figure~\ref{fig:figure1}. For regular node placements of nodes, the $A_i$ are equal and is equal to unity for all $i$; the network objective reduces to max-min fair throughput of the wireless nodes.

The node placement strategy also needs to be identified before deployment and the service region and traffic routing has to be planned well in advance.

\section{Numerical Work}
\label{sec:numerical_work}
In this section, we report the performance of the network for the different control strategies described in Section~\ref{sec:network_objective}. The plots in the section correspond to solutions of the optimization problem described in Section~\ref{sec:network_objective}, solved using Matlab.

\begin{figure}
\begin{center}
\includegraphics[scale=0.6]{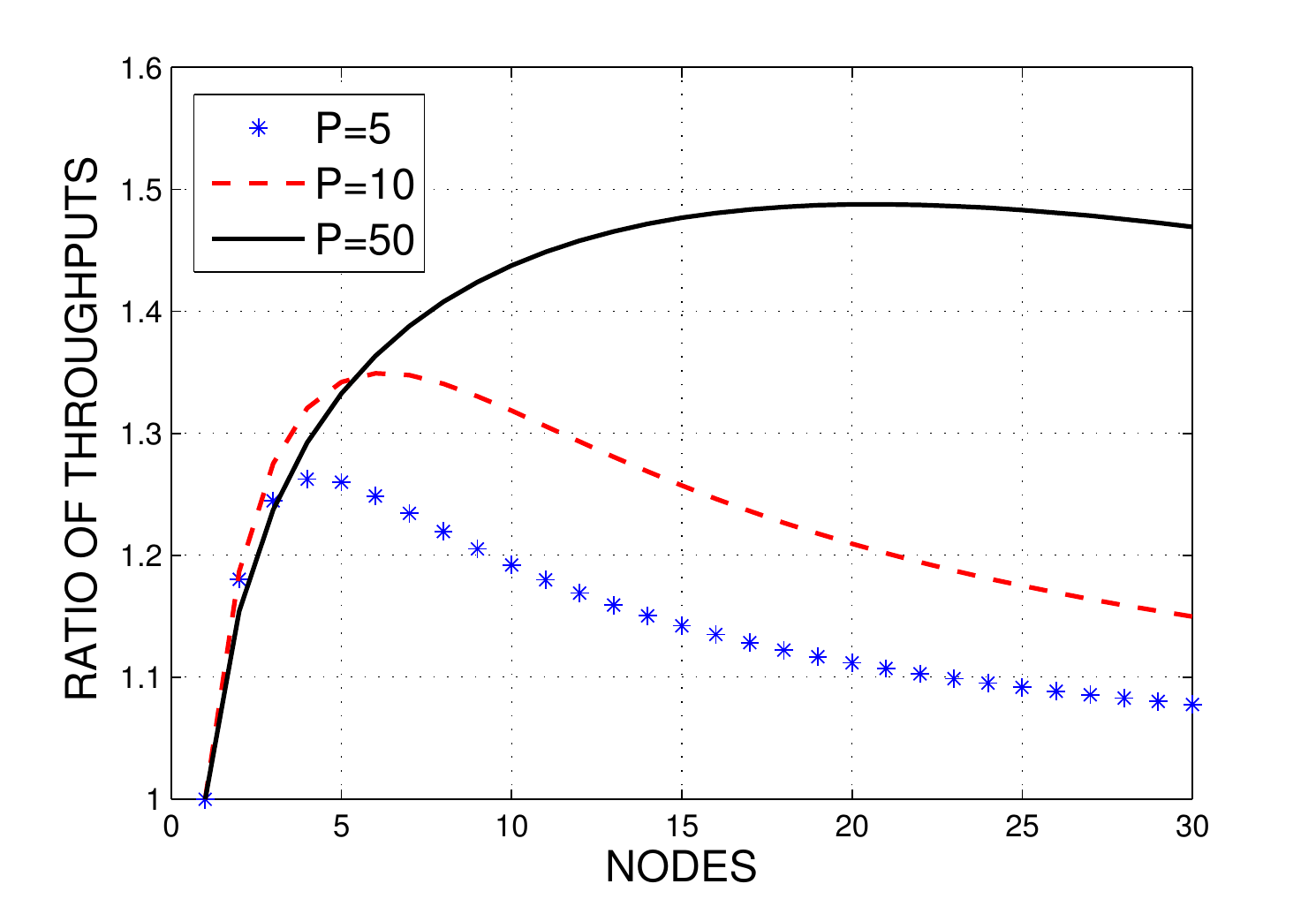}
\caption{Ratio of the max-min throughput of the time adaptation scheme and the base reference model.}
\label{fig:figure3}
\end{center}
\end{figure}
In Figure~\ref{fig:figure3}, we report the the performance of the time adaptation scheme as a function of the number of wireless nodes and the node average power $P$.
In the figure, we plot the ratio of the max-min throughput of the time adaptation scheme and the base reference model for three different values of $P$. The nodes are regularly placed with $d_1 = 1$ and $d_{i+1} = (i+1) d_1$ for all $i \geq 1$. From the figure, we infer that the
network performance with time adaptation improves for small values of $n$ for any $P$. However, we observe that the improvement in performance and the effectiveness of the time adapted system becomes negligible in comparison with the base model for large $n$ (see also equation \ref{eq:TAlimit}). This is because, as the number of nodes becomes very large, the average node transmission time for any node goes to zero.


\begin{figure}
\begin{center}
\includegraphics[scale=0.55]{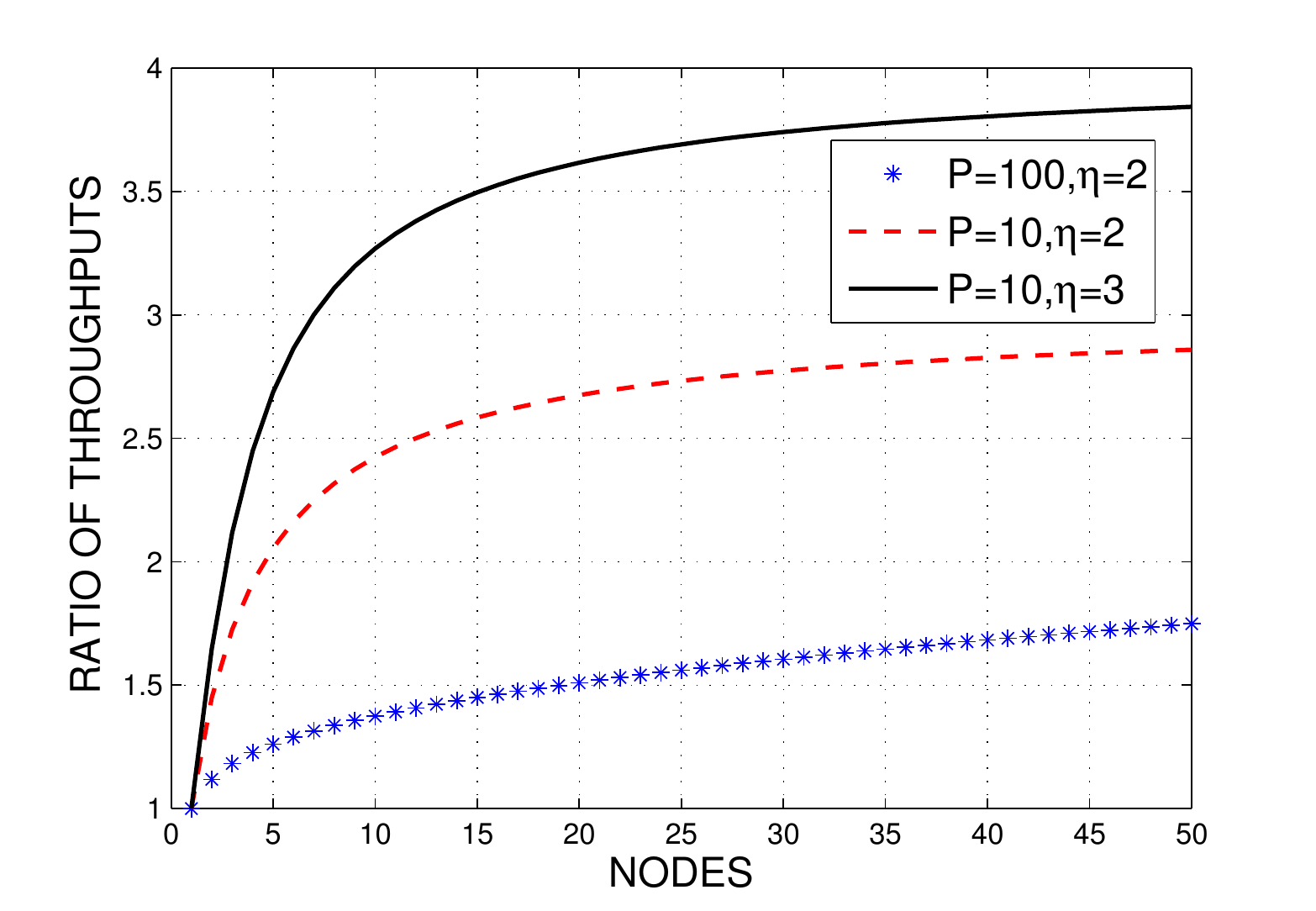}
\caption{Ratio of max-min fair throughput of power adaptation scheme and the base reference model  .}
\label{fig:figure4}
\end{center}
\end{figure}
In Figure~\ref{fig:figure4}, we plot the ratio of the max-min fair throughput of the power adaptation scheme and the base reference model as a function of the number of nodes $n$. We report the performance for different values of the node average power $P$ (with the total network power $\bar{P_n} = n P$) and path loss exponent $\eta$. The nodes are regularly placed with $d_1 = 1$ and $d_{i+1} = (i+1) d_1$ for all $i \geq 1$. From the figure, we could infer that the network performance with power adaptation improves for all $n$ and with $\eta$ (as reported in Section~\ref{sec:network_objective}).
We also note from the figure that the improvement in performance with power adaptation decreases as the node average power increases. This is due to the concave nature of the rate curve (law of diminishing returns). 

\begin{figure}
\begin{center}
\includegraphics[scale=0.6]{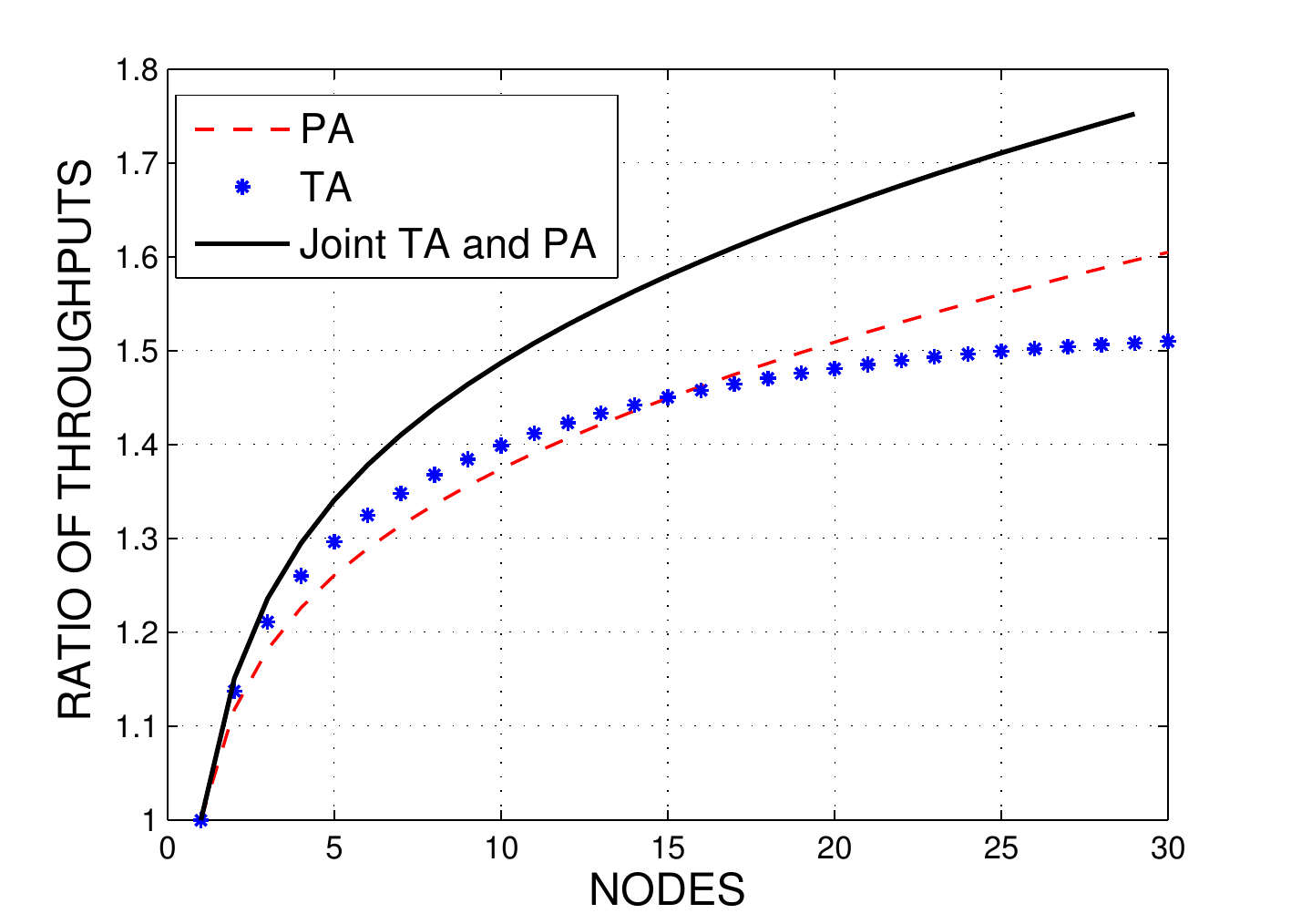}
\caption{Performance comparision of power adaptation, time adaptation and the joint time and power adaptation schemes.}
\label{fig:figure5}
\end{center}
\end{figure}
In Figure~\ref{fig:figure5}, we plot the ratio of the max-min fair throughput of the power adaptation (PA), time adaptation (TA) and joint time and power adaptation (TAPA) schemes with the base reference model.
Let the throughput of node $i$ in the joint time and power adaptation scheme be defined as $C_{i}^{TAPA}=t_i\log(1+\frac{\frac{P_i}{t_i}}{{d_i}^\eta})$, then, the max-min fair network throughput of the joint
time and power adaptation scheme is given as
\[ C^{TAPA} = \max_{\bigg\{ \substack{P_i : P_i \geq 0 \sum_{i=1}^n P_i \leq \bar{P}_n\\ t_i: t_i \geq 0, \sum_{i=0}^n t_i \leq T } \bigg\}} \min \Big(C_{1}^{TAPA}, \cdots, C_{n}^{TAPA} \Big) \]
As expected the joint time and power adaptation scheme performs better than either of the two schemes. Also,
we observe that for large number of nodes, the power adaptation scheme performs as well as the joint time and power adaptation scheme. This is because time adaptation becomes ineffective for large $n$.

\begin{figure}
\begin{center}
\includegraphics[scale=0.6]{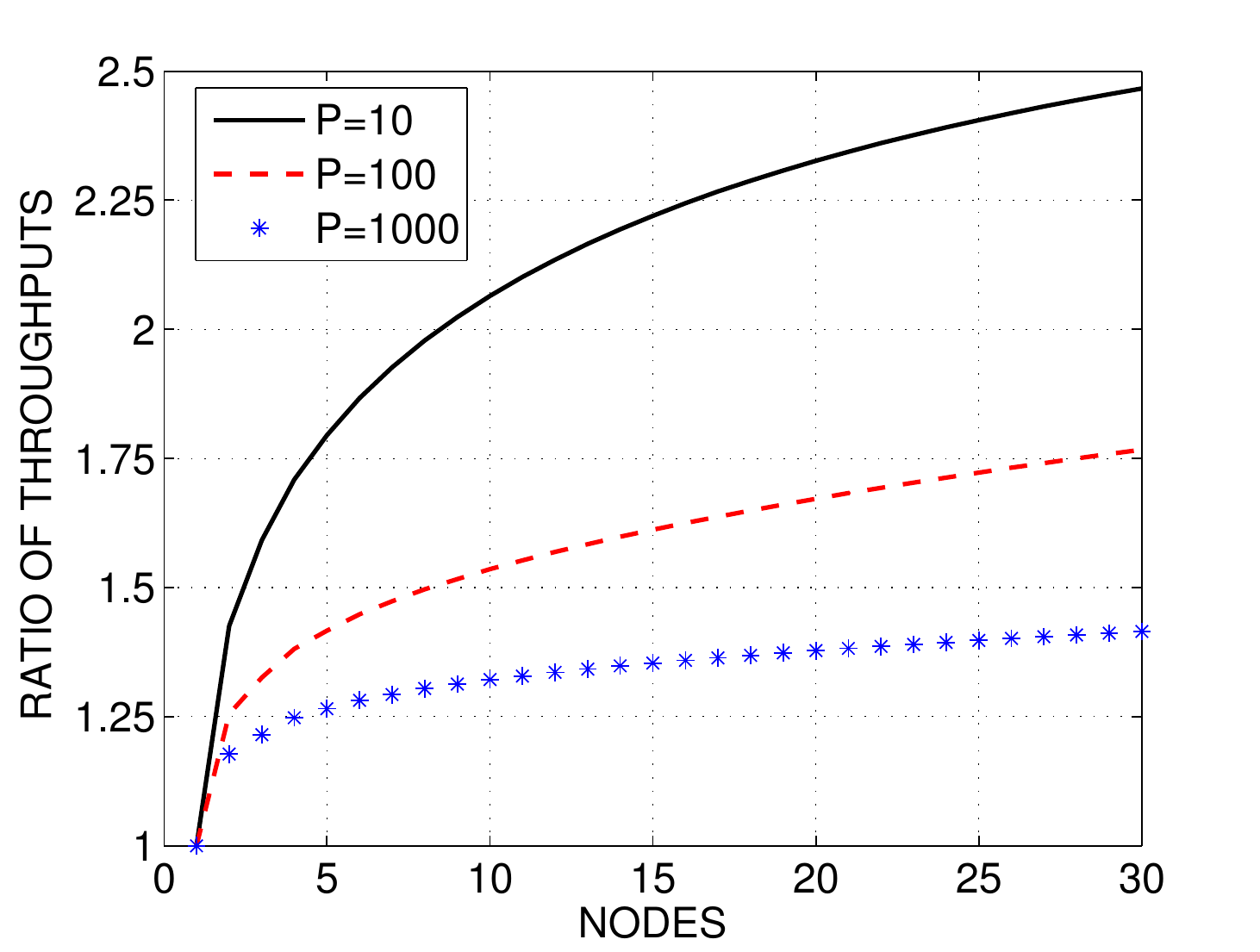}
\caption{Ratio of max-min fair throughput of the node placement scheme and the base reference model.}
\label{fig:figure6}
\end{center}
\end{figure}
In Figure~\ref{fig:figure6}, we plot the ratio of the max-min fair throughput of the node placement adaptation scheme and the base reference model. $t_i = \frac{1}{n}$ for all the nodes and $P_i = P$ for all the nodes. From the simulations, we observed that the optimal node locations are such that $n > d_1^* - 0 > d_2^* - d_1^* > \cdots > d_n^* - d_{n-1}^* > 0$ and $n > A_1^* > A_2^* > \cdots > A_n^* > 0$; the nodes farther away from the sink generate less traffic (service a smaller region) than the nodes closer to the sink. From the plot, we observe that the network performance with node placements improves for all $n$. 

\begin{figure}
\begin{center}
\includegraphics[scale=0.63]{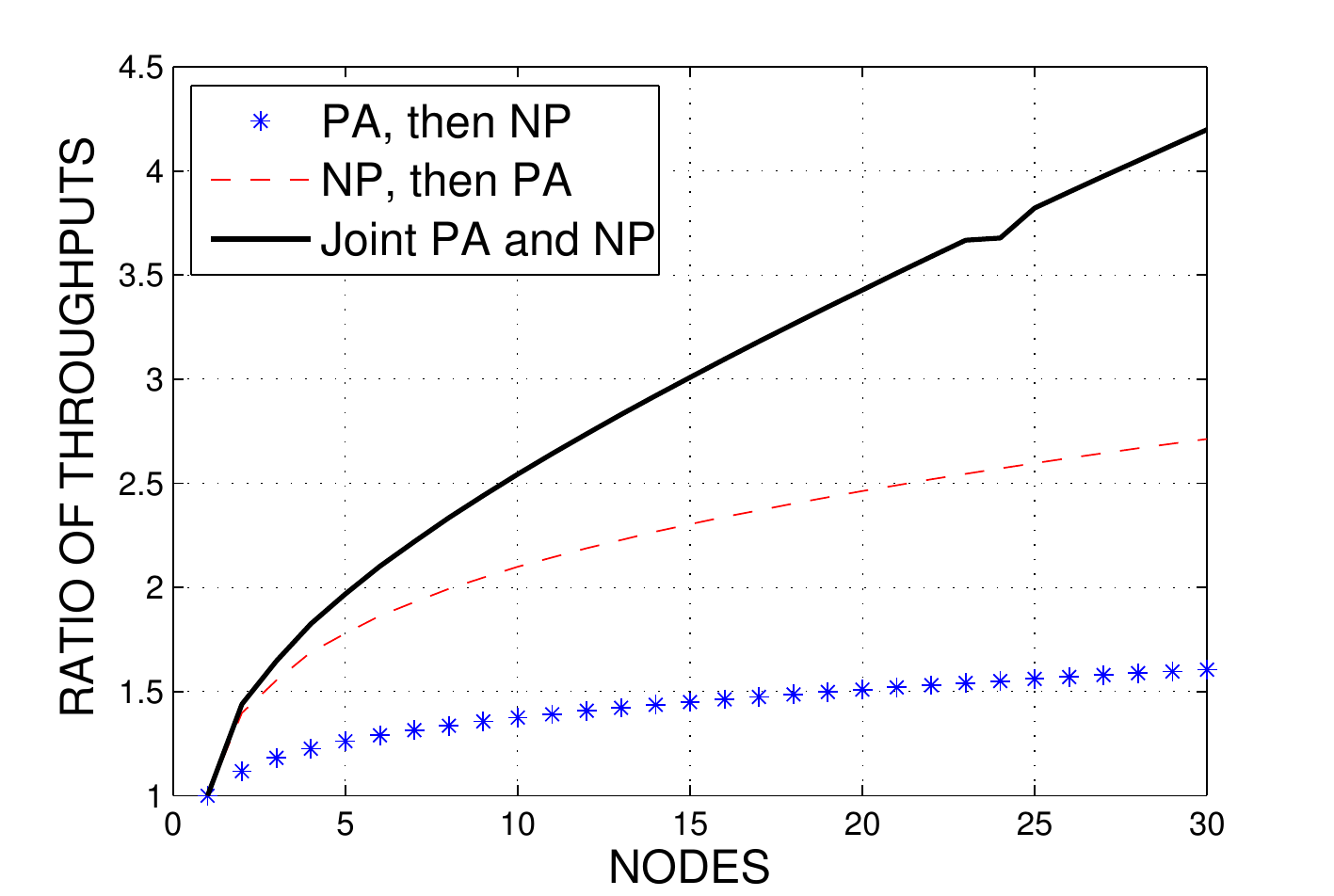}
\caption{Performance comparision of joint optimisation with other power adaptation and node placement schemes.}
\label{fig:figure7}
\end{center}
\end{figure}
In Figure~\ref{fig:figure7}, we plot and compare the performance of three different strategies involving power adaptation and node placements. For example, the scheme PA then NP aims to identify the optimal allocation for power first and then optimizes over the node placements separately. The max-min throughput of such a system is given as
\[ C^{PA,NP} = \max_{\big\{ {\{d_i\} : 0 \leq d_1 \leq \cdots \leq d_n \leq n} \big\}} \min_{i} \left( \frac{C_i(\frac{1}{n}, P_i^*,d_i)}{A_i} \right) \]
The joint power adaptation and node placement schemes aims to optimize simultaneously on the two parameters.
From the figure, we note that the joint adaptation performs better than either of the other two schemes involving PA and NP.

\section{Conclusion}
\label{sec:conclusion}
In this work, we have studied throughput optimization of a data aggregating network with simple and practical design strategies. We have restricted to single hop communication with no spatial reuse and we have studied the network performance
for different node transmission time, node transmission power and node placements.
In a data aggregating network where all the nodes generate similar traffic, the nodes that are far away from the sink suffer due to the large path loss. Most scheduling strategies are either unfair to such nodes or have poor network life time. In this paper, we have studied three simple strategies to optimize the network performance for such a data aggregating network. We have observed that time adaptation is a useful strategy for small number of users and for large $P$. Power adaptation and node placement provide better gains than time adaptation for small $P$ and for large values of $n$.
The results indicate that the network should be planned before deployment in the field for optimal performance.


In future, we would like to study the performance of the different strategies for a fading wireless channel. Also, we would like to study the benefits of multihopping without spatial reuse for the data aggregating network.

\section*{Acknowledgements}
This work was partly funded by The India-UK Advanced Technology Centre (IU-ATC) of Excellence in Next Generation Networks Systems and Services.

%

\bibliography{references.bib}
\bibliographystyle{IEEEtran}

\end{document}